\documentclass[epj,twocolumn]{webofc}
\usepackage[varg]{txfonts}   
\woctitle{XLIV International Symposium on Multiparticle Dynamics}
\def\dsigma{{\rm d} \hat\sigma}
\begin{document}
\title{Precise QCD predictions for jet production at the LHC}
%
%

\author{Jo\~{a}o Pires\inst{1,2}\fnsep\thanks{\email{joao.pires@mib.infn.it}}}

\institute{Dipartimento di Fisica, Universit\`{a} di Genova and INFN, Sezione di Genova, Via Dodecaneso 33, I-16146 Genova, Italy 
\and
 	      Dipartimento di Fisica, Universit\`{a} di Milano and INFN, Sezione di Milano, Via Celoria 16, I-20133 Milano, Italy\fnsep\thanks{address from January 2015}
          }

\abstract{%
Theoretical predictions used in experimental analysis of LHC data have an inherent theoretical uncertainty associated to the relevant order in the perturbative expansion that the observable is computed. 
In this talk we briefly introduce and motivate the topic of QCD radiative corrections stressing the impact of including higher-order QCD effects in the theory predictions and in particular we consider two-loop QCD corrections. 
We present numerical results at NNLO for gluonic jet production where jets are reconstructed using the anti-$k_T$ jet algorithm from gluon-gluon and quark-antiquark scattering and
discuss the inclusion of leading-$N_F$ quark contributions in the final-state.}
\maketitle
\section{Introduction}
\label{intro}
In proton-proton collisions, the factorised form of the inclusive cross section is given by,
\begin{equation}
{\rm d}\sigma=\sum_{i,j} \int  f_i(x_1,\mu_F^2)f_j(x_2,\mu_F^2)
 {{\rm d}\hat{\sigma}}_{ij}(\alpha_s(\mu_R^2),s/\mu_R^2,s/\mu_F^2)\;,
\end{equation}
where ${\rm d}\hat\sigma_{ij}$ is the parton-level scattering cross section for parton $i$ to scatter off parton $j$
and the sum runs over the possible parton types $i$ and $j$. The probability of finding a parton of type $i$ in the proton carrying a momentum fraction $x$ is described by the 
parton distribution function (PDF) $f_i(x)dx$.  By applying suitable cuts, one can study more exclusive observables such as the transverse momentum distribution or rapidity 
distributions of the hard objects (jets or vector bosons, higgs bosons or other new particles) produced in the hard scattering.  
The leading-order (LO) prediction is a useful guide to the rough size of the cross section, but is usually subject to large uncertainties from the 
dependence on the unphysical renormalisation and factorisation scales, as well as possible mismatches between the (theoretical) parton-level and the (experimental) hadron-level.  

\begin{figure}
\centering
\includegraphics[width=8cm,clip]{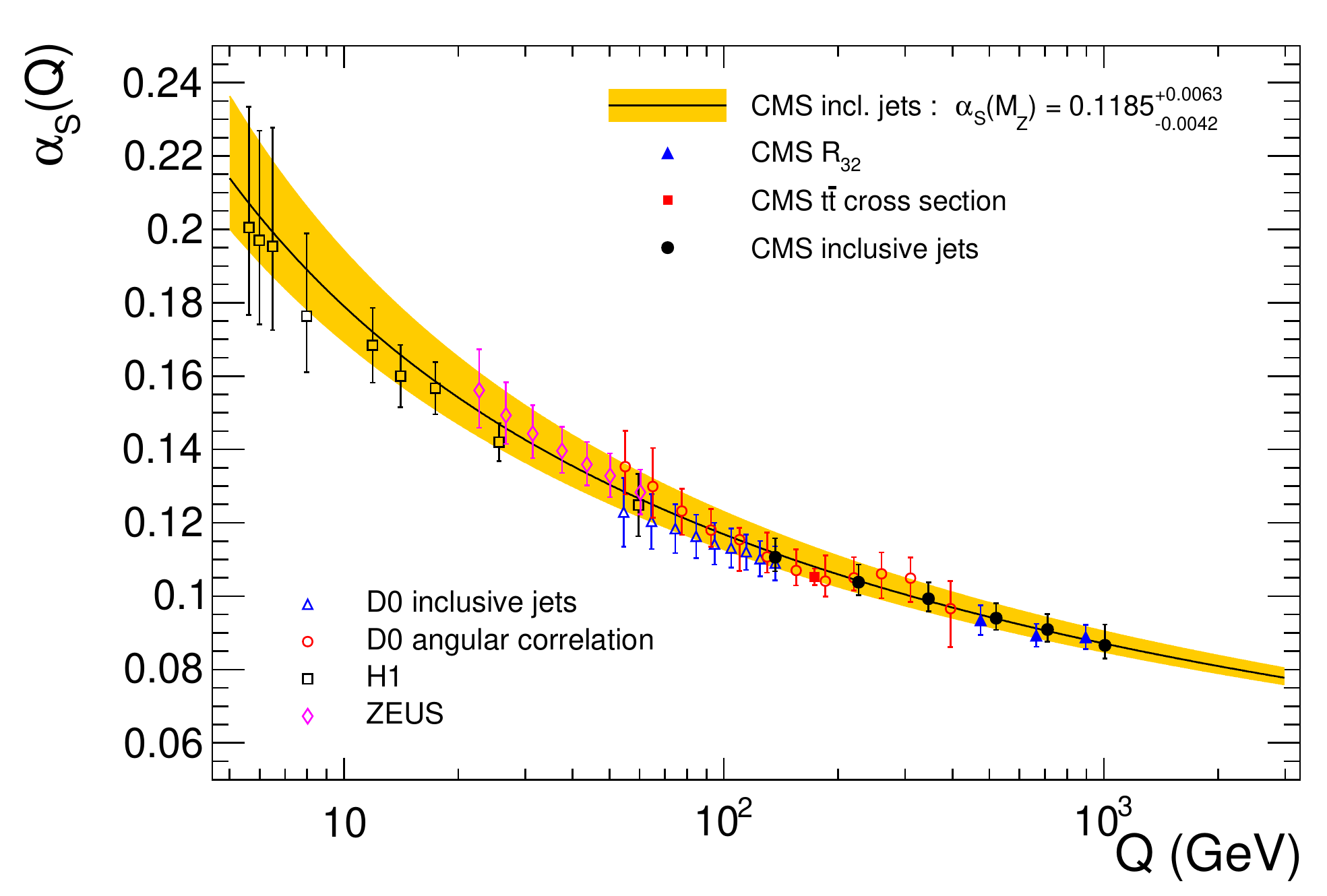}
\caption{The strong coupling $\alpha_{S}$(Q) (full line) extraction from inclusive CMS jet data and its total uncertainty (band)
using a two-loop solution to the RGE as a function of the momentum transfer Q =$p_T$ determined by the CMS collaboration in the analysis of Ref.~\cite{CMSas1}.}
\label{fig:asCMS}       
\end{figure}

\begin{figure*}
\begin{minipage}[b]{0.5\linewidth}
\centering
\includegraphics[width=8cm,angle=0]{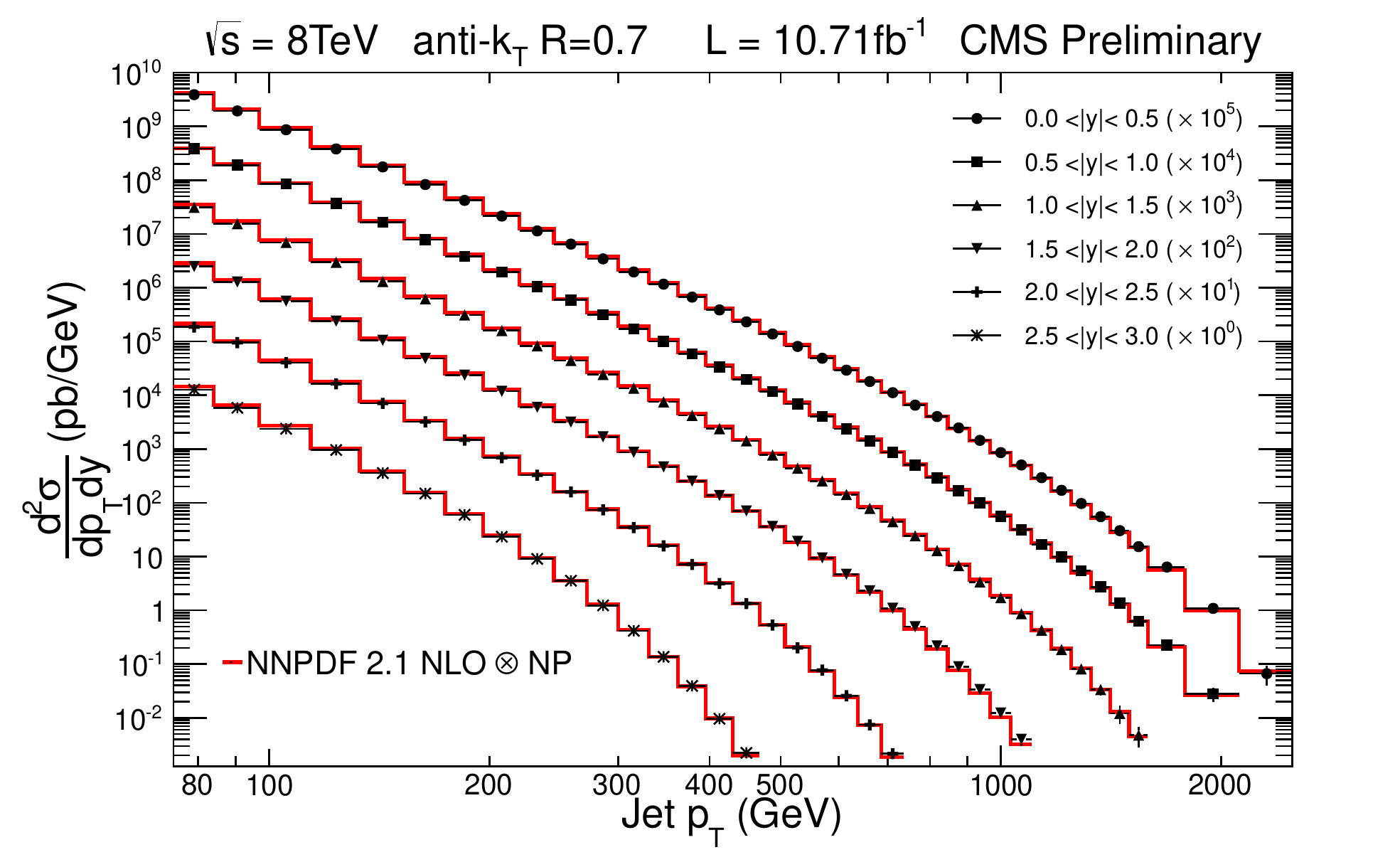}\\
\vspace{0.2cm}
(a)
\end{minipage}
\hspace{1.0cm}
\begin{minipage}[b]{0.4\linewidth}
\centering
\includegraphics[width=5.5cm]{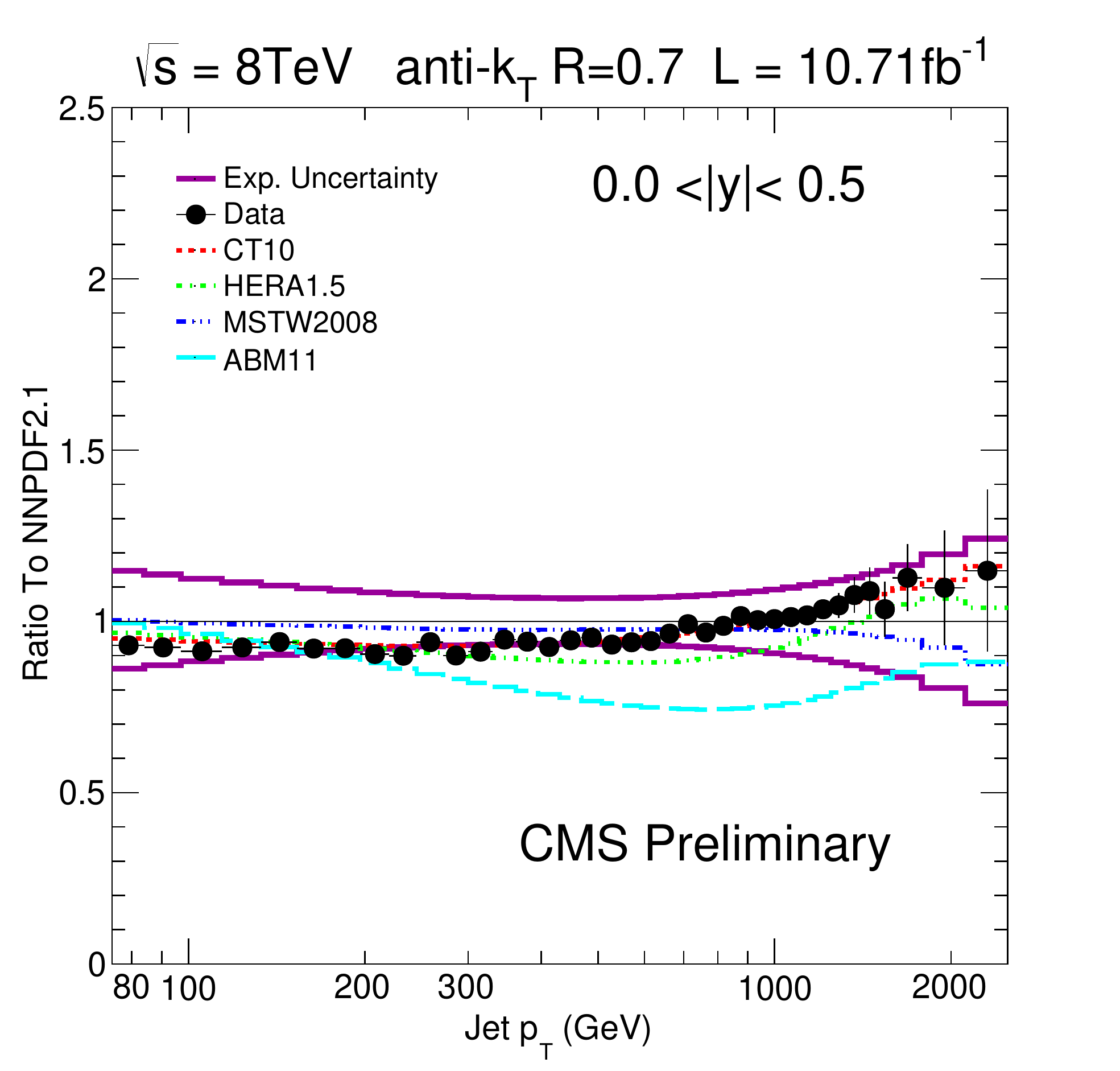}\\
\vspace{0.2cm}
(b)
\end{minipage}
\caption[]{(a) Double-differential inclusive jet cross section measurement from Ref.\cite{CMS-PAS} in comparison to NLO predictions using the NNPDF2.1 PDF set times the NP correction factor. 
(b) Ratio of data over theory at NLO times NP correction for the NNPDF2.1 PDF from Ref.~\cite{CMS-PAS}. For comparison predictions employing four other PDF sets are shown in
addition to the total experimental systematic uncertainty (band enclosed by full magenta lines). The error bars correspond to the statistical uncertainty of the data \cite{CMS-PAS}.}
\label{fig:CMSincjets}
\end{figure*}

For these reasons a description of an observable to NLO and NNLO accuracy brings several advantages. At higher order additional contributions
due to radiation from the initial and final state are included leading to a better description of transverse momentum distributions and of the modelling of final state jets. 
Moreover the inclusion of higher order effects significantly reduces the observable dependence on the choice of unphysical renormalisation and factorisation scales, therefore
reducing the theoretical uncertainty in the prediction. However, beyond the advantages mentioned above, the availability of predictions for hadronic observables at higher-order in 
perturbation theory is mandatory for observables where the perturbative expansion exhibits a slow convergence (such as the Higgs boson cross section) and also when the
achievable precision in data analysis from colliders is limited by theory.

The latter has been demonstrated in recent papers by the CMS collaboration \cite{CMSas1,CMSas2} where it is shown that the currently dominant theoretical uncertainties
limit the achievable precision in the extraction of the strong coupling constant from inclusive jet data. In the analysis using predictions from perturbative quantum chromodynamics at 
NLO, complemented with electroweak corrections, for all the six bins in $p_T$ represented by black dots in Fig.~\ref{fig:asCMS} the scale uncertainty in the theory prediction dominates 
over experimental, PDF and non-perturbative (NP) uncertainties leading to a strong coupling constant determination of $\alpha_{S}(M_Z) = 0.1185 \pm 0.0019 (exp)^{+0.0060}_{-0.0037} (theo)$.

The accuracy of this determination has only been made possible thanks to the unprecedented experimental precision of the jet measurements that allow stringent tests of QCD
to be performed. In particular, the large cross section for jet production at the LHC allows the jet measurements to be performed in multi-differential form accessing phase space regions
of $Q^2$ and $x$ not covered by previous experiments. In figure~\ref{fig:CMSincjets}(a) we show the latest measurements of the inclusive jet cross section by CMS 
with a total integrated luminosity of 10.71 $fb^{-1}$ from 8 TeV proton-proton collisions. In figure~\ref{fig:CMSincjets}(b) a more detailed comparison between data and NLO theory using
different PDF sets and non-perturbative corrections included is shown together with the systematic and statistical data uncertainties. For this measurement and across the entire $p_T$ range 
the experimental and theoretical uncertainties are roughly of the same of size. We can observe that there is an overall good agreement with the data with the different PDF sets giving identical predictions at low $p_T$.
However there is a significant mismatch in the predictions at high $p_T$ between all PDF sets and in this region the precision of the data would allow to distinguish the different sets and further
constrain the PDF's.

\begin{figure*}
\begin{minipage}[b]{0.5\linewidth}
\centering
\includegraphics[width=7.5cm,angle=0]{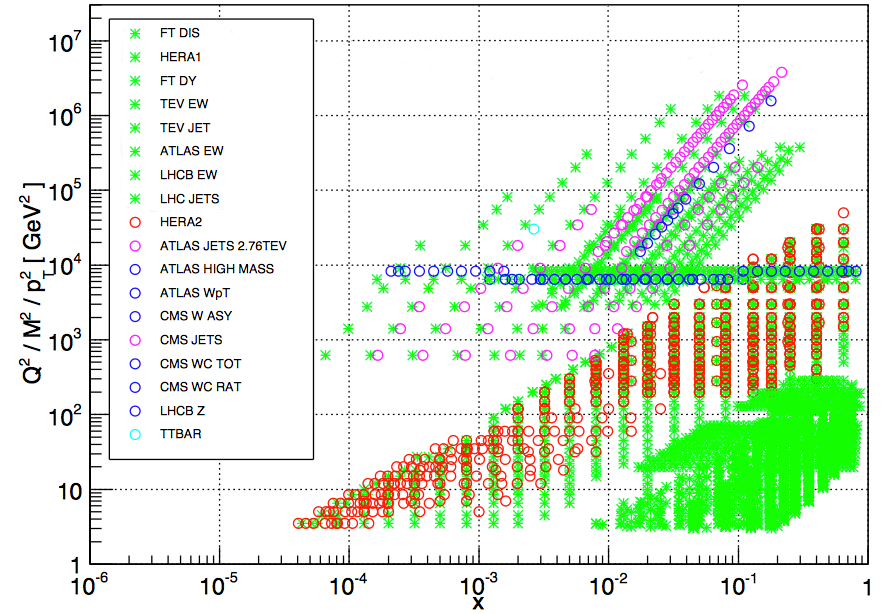}\\
\vspace{0.2cm}
(a)
\end{minipage}
\hspace{1.0cm}
\begin{minipage}[b]{0.4\linewidth}
\centering
\includegraphics[width=7.5cm]{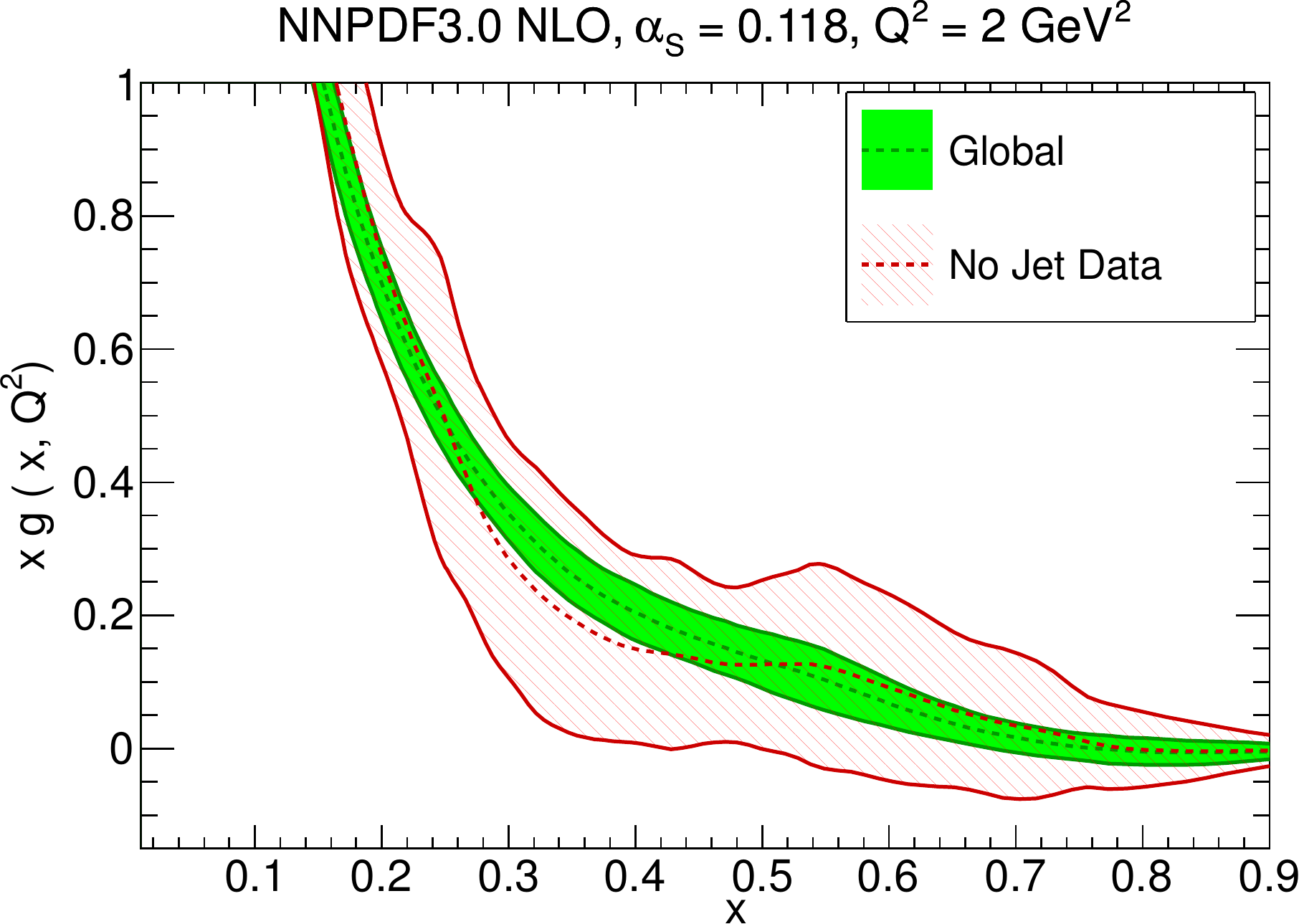}\\
\vspace{0.2cm}
(b)
\end{minipage}
\caption[]{(a) The kinematical coverage in the $(x,Q^2)$ plane of the NNPDF3.0 dataset \cite{NNPDF} where the green stars mark the data already included in NNPDF2.3, 
while the circles correspond to experiments that are novel in NNPDF3.0. (b) Comparison of the gluon in a fit to a dataset without jet data and in the global fit at NLO \cite{NNPDF}.}
\label{fig:NNPDF30}
\end{figure*}

The influence of the jet data on parton distribution functions is also discussed in \cite{CMSas1}. Here we review the analysis performed by the NNPDF collaboration in \cite{NNPDF} which
presented recently the NNPDF3.0 set in Ref~\cite{NNPDF}. The full set of experimental datapoints included in this set is shown in figure~\ref{fig:NNPDF30} (a).
In particular, it contains inclusive jet data from ATLAS and CMS, namely the inclusive jet production measurements at 7 TeV and also the ATLAS data at $\sqrt{s}$=2.76 TeV,
with the goal of assessing the gluon distribution in the proton at medium to large values of the momentum fraction $x$. This is shown in~\ref{fig:NNPDF30} (b) where the NLO
gluon PDF and its uncertainty obtained from a global fit is shown in green while the fit plotted in red is the resulting fit if all jet data are removed from the global dataset. 
It is observed~\cite{NNPDF} that a significant reduction of the gluon uncertainty and a reliable determination of the gluon is achieved by keeping the jet data in the dataset. For this reason and
with the aim of making jet data consistently included in NNLO PDF fits, the perturbative theory calculation should be avaliable at the same order.  
 
In order to achieve these cutting edge extractions from the LHC jet data, experimental and theory collaborations compare their measurements to fixed-order calculations provided 
by Monte Carlo event generators that encode the predictions of QCD. For jet cross sections beyond NLO the computations are done in the form of a parton-level generator, 
which is a numerical program, providing full kinematical information on parton-level final states to a given perturbative order. Within the parton-level generator, the probability
for a specific final state to occur is computed by weighting the generated phase space point with all scattering matrix elements relevant to the final state under consideration.
 
\section{NNLO calculations}
\label{sec-1}
Calculations of NNLO corrections have a significant increase in their complexity. An $n$-jet observable requires two-loop $n$-particle matrix elements (double-virtual correction), 
one-loop ($n$+1)-particle matrix elements (real-virtual correction) and tree-level ($n$+2)-particle matrix elements (the double-real correction) which contain IR singularities due
to one or two particles becoming unresolved (soft and collinear) in tree- level and one-loop matrix elements. 

For this reason, successful computations with exact NNLO accuracy
in fully differential form for hadron collider observables have been achieved only for very specific processes, with the bottleneck for tackling the real-radiation contribution of a generic process
being the absence of a general method to extract implicit infrared (IR) poles from real-virtual and double-real contributions in a fully differential way, i.e., without doing the phase space integration.
However, enormous progress has been achieved in this direction thanks to extensive work on the development of subtraction schemes at NNLO, with the most recent published results being obtained
using the frameworks of antenna subtraction \cite{antennaFF}, sector decomposition~\cite{SD}, $q_T$ subtraction~\cite{qT}, and sector improved residue subtraction~\cite{stripper}, which we briefly review. 

The sector decomposition method involves a split of the matrix elements and the final state phase space into phase space sectors and the cancellation of IR singularities is achieved by numerically integrating them. 
It has been applied to $2\to1$ processes (such as $pp\to H$~\cite{sdH}, $pp\to V$~\cite{sdV}). The $q_T$ method is restricted to processes with colourless final states ($pp\to H,V,VH,VV$~\cite{qTresum}) 
and exploits the universality of the resummation of large logarithmic corrections at small transverse momentum to regulate the divergences at NNLO in the small $q_T$ limit. 
Finally the numerical sector improved residue subtraction approach combines an initial partitioning of the phase space (inspired by the NLO FKS approach~\cite{FKS}) with sector 
decomposition to generate sectors at NNLO where the universal singular structure of the divergences of the process are explicitly isolated and are then subtracted using their known explicit form. 
It has been applied to single top~\cite{singleT} production and to the total and differential cross section for $t\bar{t}$ quark production~\cite{ttb} and to $pp\to H+j$~\cite{Hj}. 
More recently NNLO QCD corrections for the hadroproduction of a pair of off-shell photons in the limit of a large number of quark flavours were obtained in~\cite{VVoff} using a
dedicated fully factorised parametrisation for the phase space for the RR and RV contributions based on an extension of the proposal in~\cite{nonlinearmaps}.

In this talk I will discuss the antenna subtraction approach proposed in \cite{antennaFF} for colourless initial states. In this approach the computation of the NNLO coefficient is organised according to three 
integration channels each identified by the multiplicity of the final state,

\begin{eqnarray}
\dsigma_{ij,NNLO}&=&\int_{{\rm{d}}\Phi_{m+2}}\left[\dsigma_{ij,NNLO}^{RR}-\dsigma_{ij,NNLO}^S\right]
\nonumber \\
&+& \int_{{\rm{d}}\Phi_{m+1}}
\left[
\dsigma_{ij,NNLO}^{RV}-\dsigma_{ij,NNLO}^{T}
\right] \nonumber \\
&+&\int_{{\rm{d}}\Phi_{m}}\left[
\dsigma_{ij,NNLO}^{VV}-\dsigma_{ij,NNLO}^{U}\right].
\label{eq:NNLOantsub}
\end{eqnarray}

For each choice of initial state partons $i$ and $j$, each of the square brackets is finite and well behaved
in the infrared singular regions. In particular all physical IR singularities in the double real contribution and real-virtual are subtracted from the matrix elements
by the contributions $\dsigma_{ij,NNLO}^{S,T}$ which have the property that they reproduce the singularities of the matrix elements of both contributions.
By observing that the structure of the divergences in QCD matrix elements has a process independent universal form, single unresolved emission (NLO) and double-unresolved emission (NNLO) 
can be described using antenna functions that account for all known possible unresolved configurations and their overlap. The name antenna comes from organising the subtraction terms using 
hard particle pairs (emitters) with unresolved colour ordered particle radiation emitted in-between. In this way when the numerical integration of each channel is performed over the singular 
regions of soft and collinear emission the Monte Carlo integral is now regulated.
The antenna functions then depend only on the momenta of the emitters and the unresolved
particles and are thus sufficiently simple to allow analytic integration of the counterterms $\dsigma_{ij,NNLO}^{S,T}$. 
For hadron collider observables the required set of antennae needed includes contributions due to radiative corrections from partons in the initial state~\cite{IIant}.
It is a feature of the antenna subtraction method that the explicit $\epsilon$-poles in the dimensional regularization parameter of one- and two-loop matrix elements are cancelled 
analytically and locally against the $\epsilon$-poles of the integrated antenna subtraction terms. For hadron collider observables this was achieved in the context of the dijet calculation, in particular, for
gluonic jet production in $gg\to$ jets at leading colour~\cite{ggLC} and at sub-leading colour~\cite{ggSLC} and for the leading colour calculation of $q\bar{q}\to$ jets in~\cite{qqb}.
In Ref.~\cite{hjant} precise predictions in the $gg$-channel for Standard Model Higgs boson production in association with a hadronic jet at NNLO in QCD have also been derived while the
generalisation of the method for the case of heavy quark production at NNLO was developed in Refs.~\cite{ttbant}, and first numerical results of this approach have been presented in Ref.~\cite{ttbnum}. 

For the remainder of this contribution we report on the calculation of the next-to-next-to-leading order (NNLO) QCD corrections to jet production at hadron colliders. 
\begin{figure}
\centering
\includegraphics[width=8cm,clip]{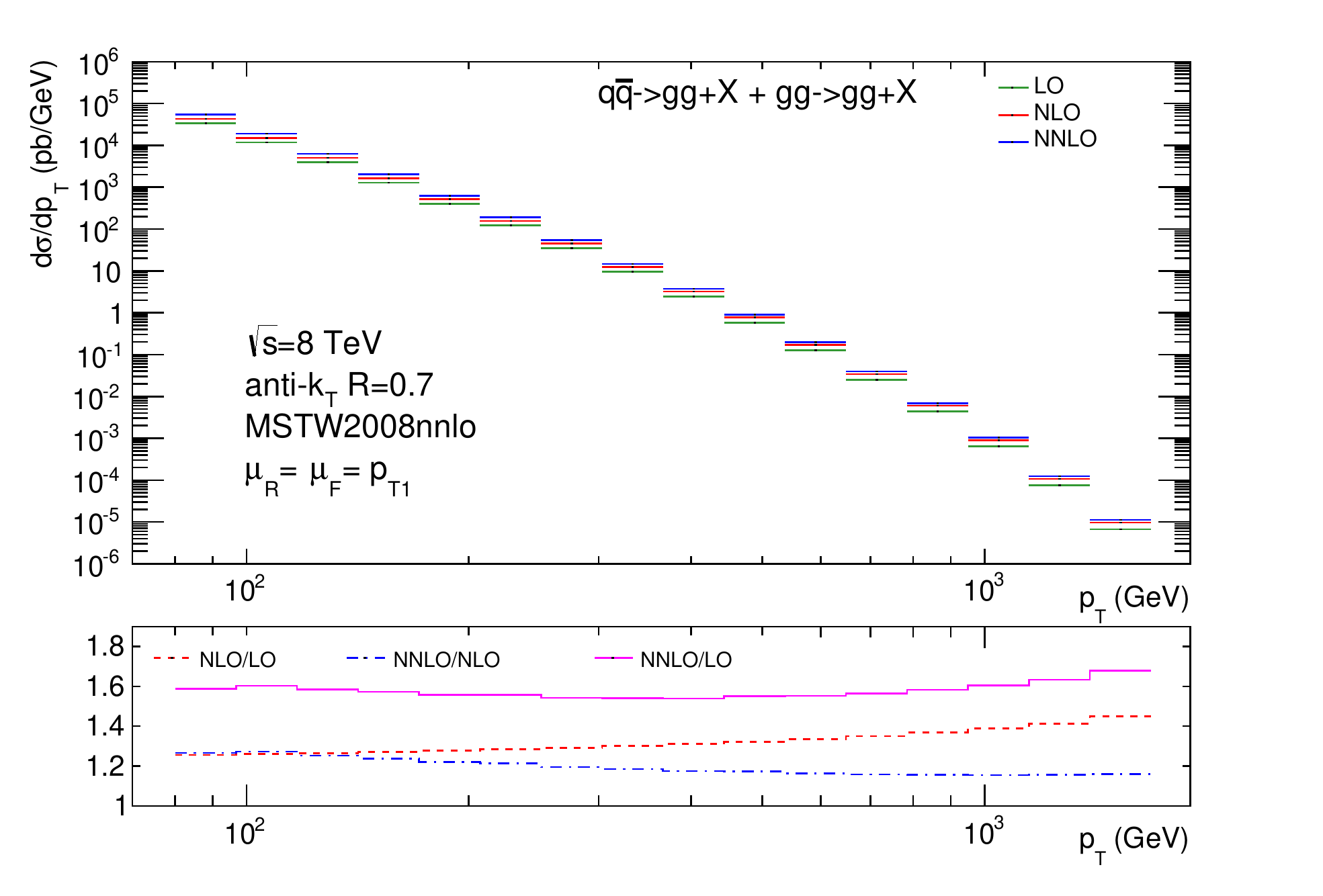}
\caption{Inclusive jet transverse energy distribution, $d\sigma/dp_T$, for jets constructed with the anti-$k_T$ algorithm with $R=0.7$ and with $p_T > 80$~GeV, $|y| < 4.4$ and $\sqrt{s} = 8$~TeV at NNLO (blue), NLO (red) and LO (dark-green). The lower panel shows the
ratios of different perturbative orders, NLO/LO, NNLO/LO and NNLO/NLO~\cite{LL2014}.}.
\label{fig:sjinc}       
\end{figure}

\begin{figure*}
\begin{minipage}[b]{0.5\linewidth}
\centering
\includegraphics[width=7.0cm,angle=0]{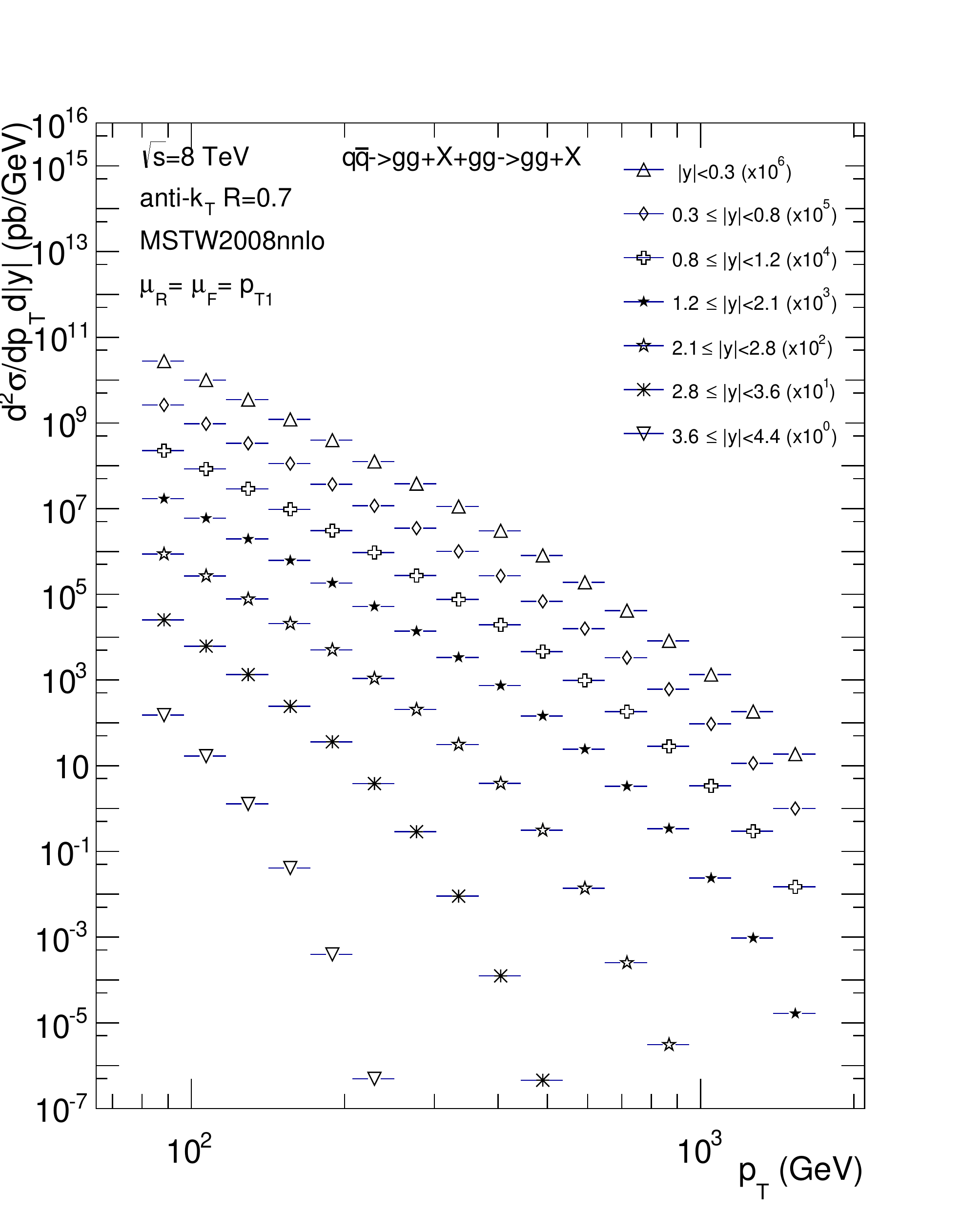}\\
\vspace{0.2cm}
(a)
\end{minipage}
\hspace{0.5cm}
\begin{minipage}[b]{0.5\linewidth}
\centering
\includegraphics[width=7.0cm,angle=0]{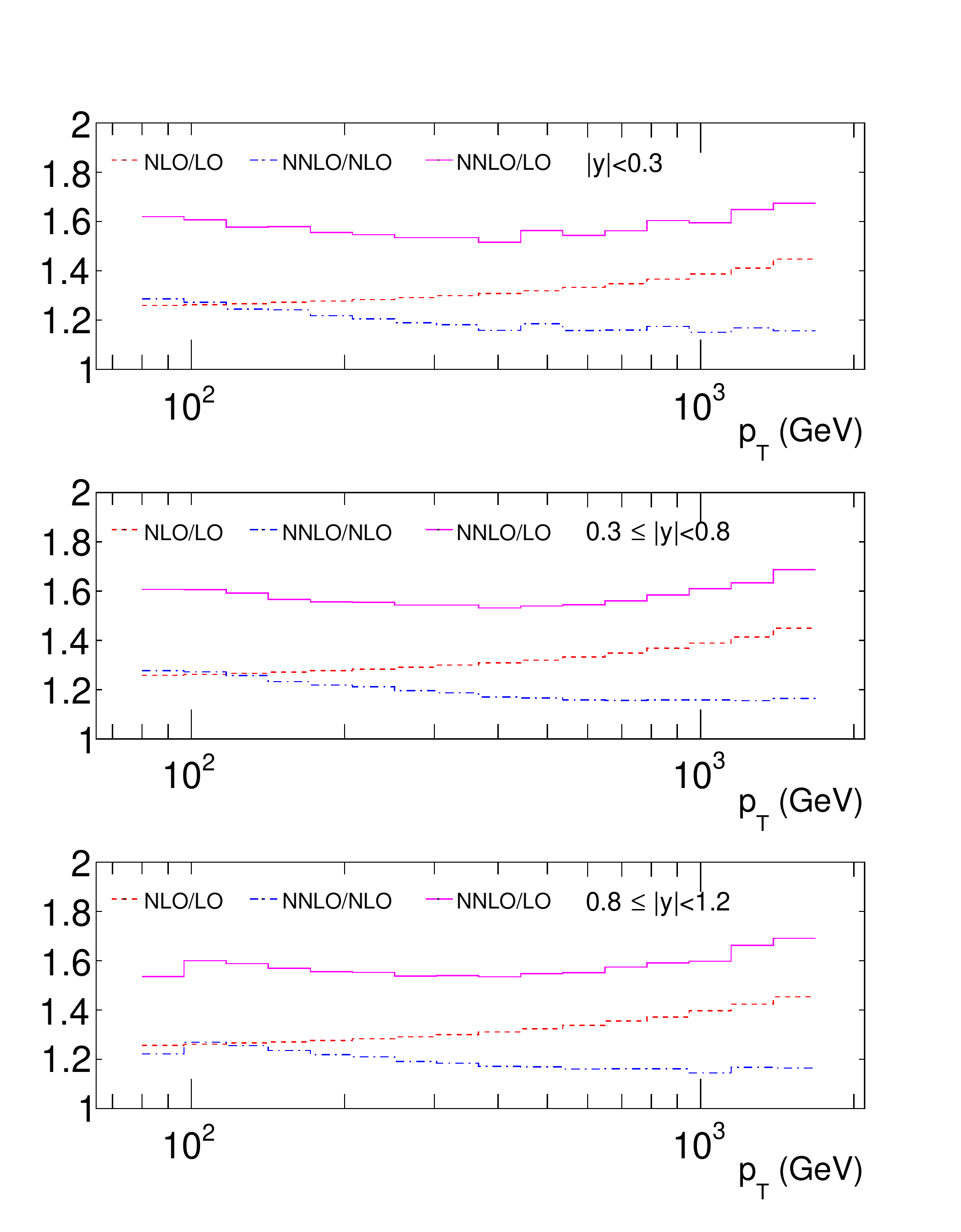}\\
\vspace{0.2cm}
(b)
\end{minipage}
\caption[]{(a) The doubly differential inclusive jet transverse energy distribution, $d^2\sigma/dp_T d|y|$, at $\sqrt{s} = 8$~TeV for the anti-$k_T$ algorithm with $R=0.7$ and for $p_T > 80$~GeV and various $|y|$ slices and (b) double differential $k$-factors for $p_T > 80$~GeV and three $|y|$ slices: $|y | < 0.3$, $0.3 < |y| < 0.8$ and $0.8 < |y| < 1.2$~\cite{LL2014}.}
\label{fig:ddifsjinc}
\end{figure*}

\section{Numerical results}
In this section we present our numerical results obtained with the parton-level generator NNLOJET for fully differential dijet production in NNLO QCD.
\subsection{Gluonic jet production at NNLO}
\label{sec-2}
Our numerical studies are derived for proton-proton collisions at the centre-of-mass energy $\sqrt{s}=8$ TeV and contain the NNLO calculation of
the full colour $gg\to$ gluons contribution combined with the leading colour $q\bar{q}\to$ gluons. These contributions have been computed in Refs.~\cite{ggLC,ggSLC,qqb}
and we present here the single jet inclusive cross section (where every identified jet in an event that passes the selection cuts contributes, such that a single event potentially enters the distributions multiple times).

Jets are identified using the anti-$k_{T}$ algorithm with resolution parameter R = 0.7 and ordered in transverse momentum. An event is retained if 
the leading jet has $p_{T1}$ > 80 GeV and rapidity $|y|$< 4.4. For all predictions we used the MSTW08NNLO distribution functions~\cite{MSTW}, including the evaluation of the LO and NLO contributions,
with the value of $\alpha_s$ provided by the PDF set through the LHAPDF~\cite{lhapdf} interface. Factorization and renormalization scales ($\mu_{F}$ and $\mu_{R}$) are chosen dynamically on an event-by-event basis. 
As default value, we set $\mu_F=\mu_R=\mu$ and set $\mu$ equal to the transverse momentum of the leading jet so that $\mu = p_{T1}$. 

Figure~\ref{fig:sjinc} shows the single jet inclusive cross section as a function of the jet $p_T$ at LO, NLO and NNLO, for the central scale choice $\mu_R=\mu_F=p_{T1}$ from Ref.~\cite{LL2014}.
The NNLO/NLO $k$-factor in the lower panel shows the size of the higher order NNLO effect to the cross section in each bin with respect to the NLO calculation. 
For this scale choice we see that the NNLO/NLO k-factor is approximately flat across the $p_T$ range corresponding to a 27-16\% increase compared to the NLO cross section. 
Note that in the combination of $q\bar{q}\to gg + gg\to gg$ channels, the gluon-gluon initiated channel dominates. The NNLO/NLO $k$-factor for the $q\bar{q}\to gg$ channel alone is roughly 5\%.

Figure~\ref{fig:ddifsjinc} we show the inclusive jet cross section in double-differential form in jet $p_T$ and rapidity bins at NNLO from Ref.~\cite{LL2014}. The $p_T$ range is divided into 16 jet-$p_T$ bins and seven rapidity
intervals over the range 0.0-4.4 covering central and forward jets. The double-differential $k$-factors for the distribution in figure~\ref{fig:ddifsjinc} (a) for three rapidity slices: |y| < 0.3, 0.3 < |y| < 0.8 and 0.8 < |y| < 1.2 are shown
in figure~\ref{fig:ddifsjinc} (b). We observe that the NNLO correction increases the cross section between 27\% at low $p_T$ to 16\% at high $p_T$ with respect to the NLO calculation (blue dot-dashed line) 
and this behaviour is similar for all three rapidity slices.

We note that as discussed in reference~\cite{SC} all predictions shown in figures~\ref{fig:sjinc} and \ref{fig:ddifsjinc} can also be obtained by evaluating the single jet inclusive cross section 
using as a scale choice $\mu_R=\mu_{F}=\mu=p_T$ where in this case each jet in every event is binned with the weight evaluated at the scale ${p_T}$ of the jet. While at LO the two final state partons
generate two jets with equal transverse momentum $p_{T1}=p_{T2}=p_T$ and the two scale choices coincide, radiative corrections can generate subleading jets and the effects of the 
different scale choice in the theory prediction become apparent at NLO and NNLO. As it was shown in~\cite{SC} for jet $p_T$'s in the range of 100 GeV$\sim$1500 GeV the effect
of the different scale choice on the absolute NNLO cross section is at the level of 5\% and decreasing at high $p_T$.

\subsection{Leading-$N_F$ contributions at NNLO}
\begin{figure}
\centering
\includegraphics[width=8cm,clip]{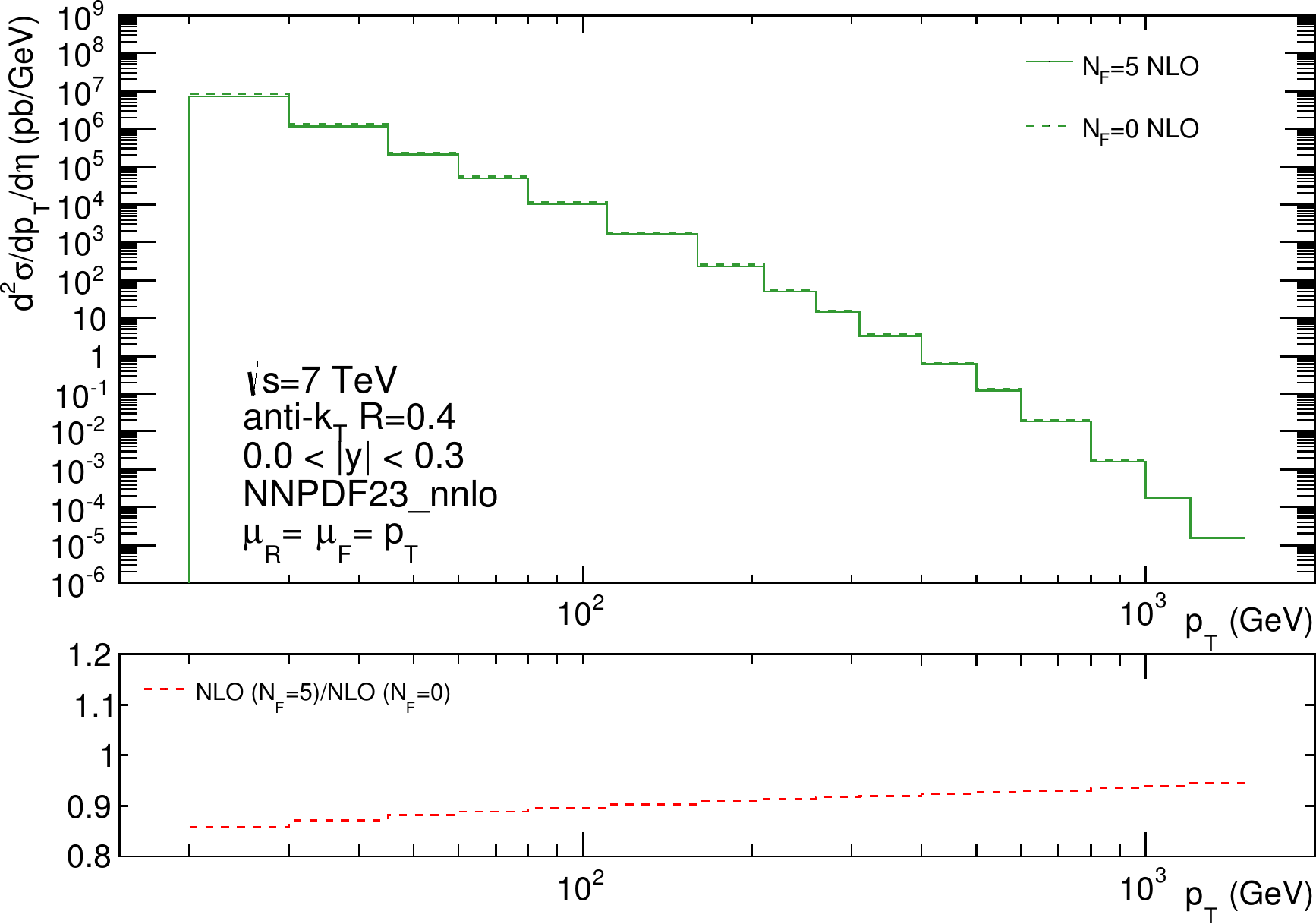}
\caption{NLO $gg$-channel predictions evaluated with five and zero quark-flavours.}
\label{fig:NLONF}       
\end{figure}

In this section we discuss the inclusion of the leading-$N_F$ quark contributions in the final-state.
From the knowledge of the full $gg$-channel cross section at NLO (plotted in figure~\ref{fig:NLONF}) we observe that the inclusion of the $N_F$ contributions leads to
a reduction of the $gg$-channel cross section between 5 and 15\%, with a larger effect at low-$p_T$. 
Motivated by the fact that at low-$p_T$ the $gg$-channel is the dominant channel for jet production, we consider in this section the calculation of its leading-$N_F$ contributions at NNLO.
The relevant partonic processes of the NNLO coefficient are listed in table~\ref{tab-1}.

For the calculation we apply the antenna subtraction method, following closely the computation in the dijet final state of the NNLO corrections in the $gg$ and $q\bar{q}$ 
channels. In this way we obtained the relevant double-real and real-virtual ${\rm d}\sigma_{NNLO}^{S}$ and
${\rm d}\sigma_{NNLO}^{T}$ subtraction terms that render the leading-$N_F$ double-real and real-virtual emission partonic processes 
finite in their single and double unresolved limits and their overlap.

In comparison with the pure gluons only calculation we needed to consider $27$ unresolved limits as opposed to $11$ and for this reason we do not provide here the explicit expressions of all the limits.
The increase in the number of unresolved singular limits that need to be subtracted is explained by the fact that for both single and double unresolved configurations the NNLO double-real partonic contribution collapses
to low-multiplicity processes with either a $q\bar{q}$ pair in the final state or a pure gluon final state. Moreover, initial-state collinear singularities generate partonic contributions which
collapse into different hard initial state processes such that only after carefully combining all the mass-factorisation relevant pieces the NNLO leading-$N_F$ contribution is infrared finite.

In order to cross check the validity of our results we implemented a serious of numerical tests for both the double-real and real-virtual contributions. In both cases we generated 
a series of phase space points that approach a given double or single unresolved limit and for each generated point we compute the ratios

\begin{equation}
R=\frac{{\rm d}\sigma_{NNLO}^{RR}}{{\rm d}\sigma_{NNLO}^{S}} \qquad R=\frac{{\rm d}\sigma_{NNLO}^{RV}}{{\rm d}\sigma_{NNLO}^{T}}\;,\nonumber
\end{equation}
between the relevant matrix element of the process and the antenna contributions. These ratios should approach unity as we get closer to any singularity and the quality of this convergence represents
the first check on the implementation of the NNLO antenna subtraction method. Moreover we will integrate both contributions over the phase space regions corresponding to 
soft and collinear emission provided that precisely 2-jets are observed by the jet algorithm. With this second test we check the convergence of the numerical integration and that
the results obtained are, for small values of the cut-off parameter, independent of the phase space generation cut-off implemented by the phase space generator. 

\begin{figure*}
\begin{minipage}[b]{0.3\linewidth}
\centering
\includegraphics[width=5.0cm,angle=0]{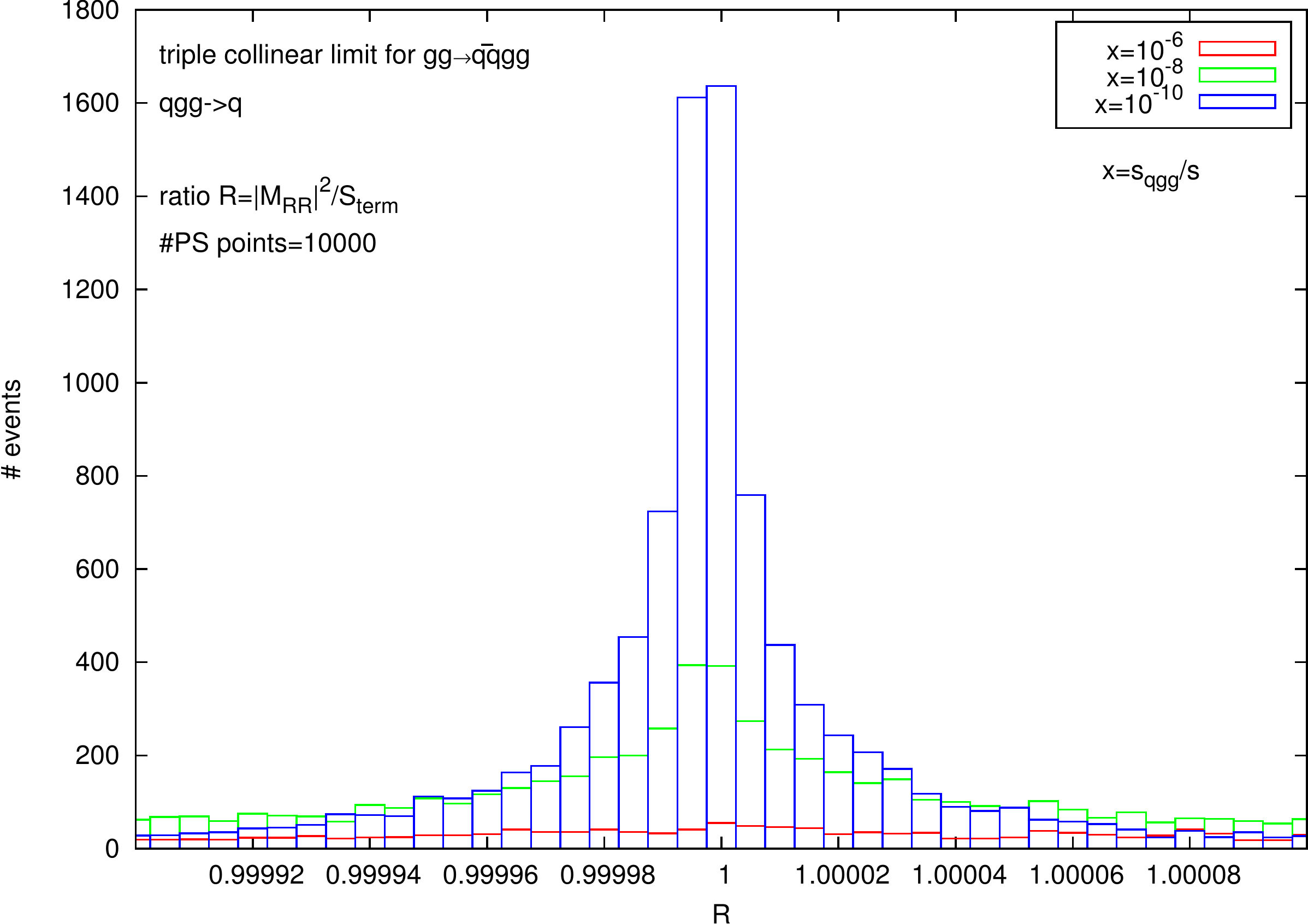}\\
\vspace{0.2cm}
(a)
\end{minipage}
\hspace{0.5cm}
\begin{minipage}[b]{0.3\linewidth}
\centering
\includegraphics[width=5.0cm,angle=0]{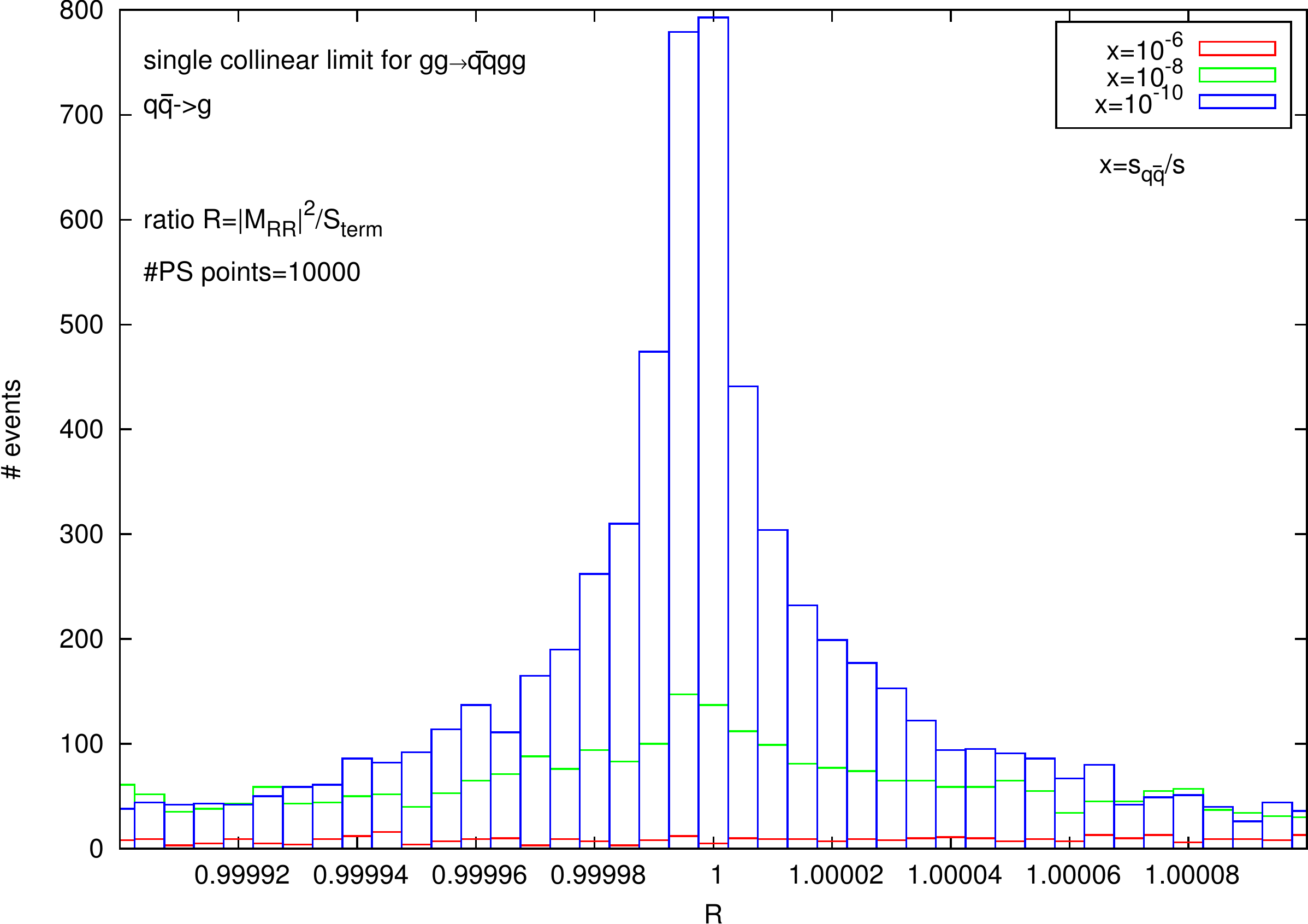}\\
\vspace{0.2cm}
(b)
\end{minipage}
\hspace{0.5cm}
\begin{minipage}[b]{0.3\linewidth}
\centering
\includegraphics[width=5.0cm]{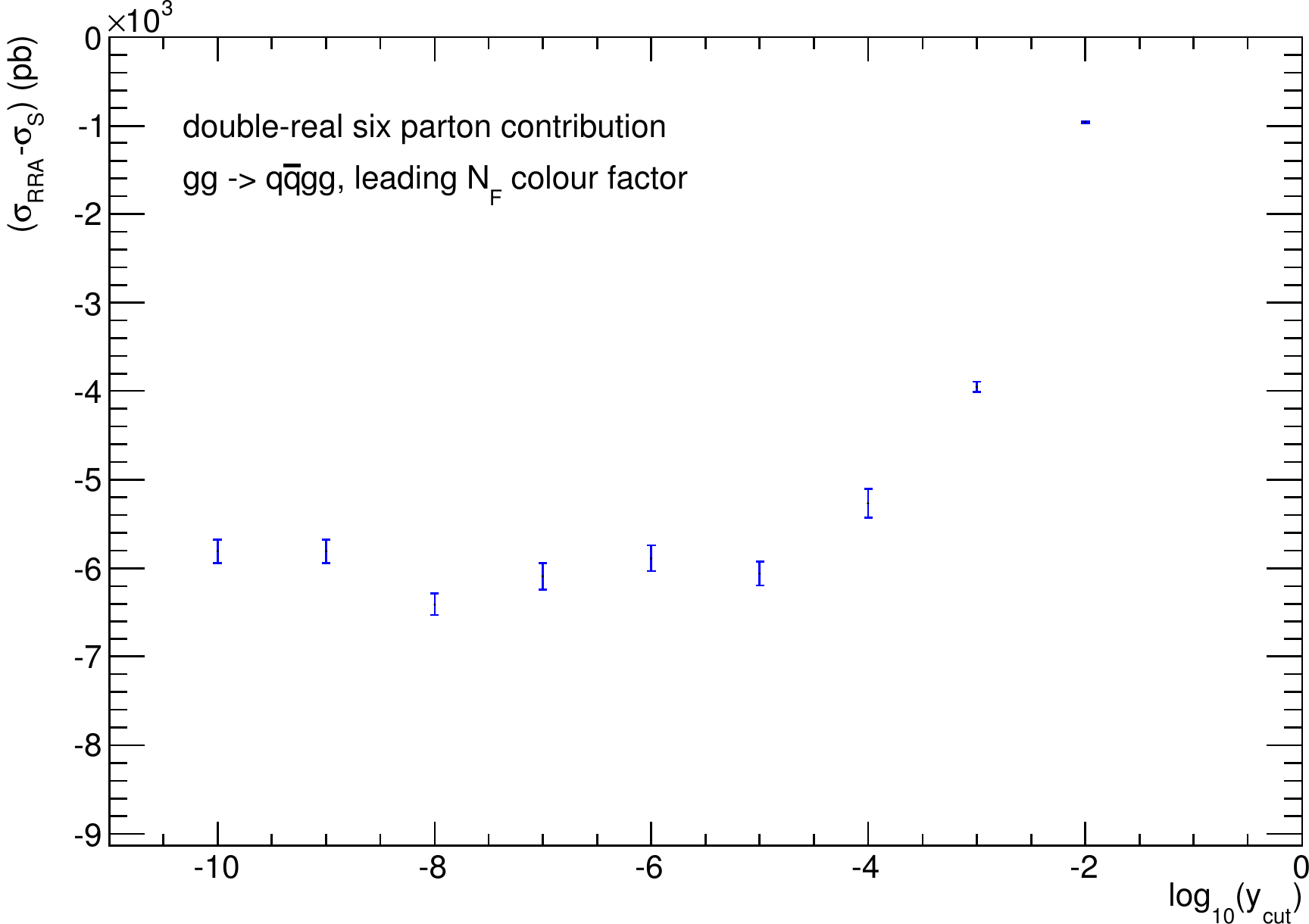}\\
\vspace{0.2cm}
(c)
\end{minipage}
\caption[]{Distribution of $R$ for 10000 phase space points for (a) triple collinear limit $qgg\to q$ configurations and (b) single collinear $q\bar{q}\to g$ configurations. For both cases the value of $x$ defined on the plot
controls the approach to each singular configuration. On the right (c) the numerical values for the integration of the subtracted double-real emission contribution as a function of the phase space
generation cut-off parameter $y_{cut}$.}
\label{fig:RRint}
\end{figure*}

As an example for the double real process
we show in figure~\ref{fig:RRint} the generation of 10000 random phase space points for both a triple collinear (a) and a single collinear configuration (b). For the single collinear case we show
the configuration where a gluon splits into a collinear $q\bar{q}$ pair and in this case we compute the ratio between the matrix element and the antenna subtraction term, by combining two phase space points
related to each other by a rotation of the system of unresolved partons by an angle of $\pi/2$ around the resultant parton direction. We observe that this is sufficient to subtract the angular correlations
in the $g\to q\bar{q}$ splittings with the subtraction terms constructed from azimuthally averaged antenna functions. In this way, for any infrared-safe 2-jet observable, the angular correlations vanish 
after integration over the azimuthal angle and in particular are not relevant for reproducing the correct $1/\epsilon$-poles in the virtual contributions. 
To reinforce this statement, Fig.\ref{fig:RRint} (c) shows the convergence 
of the numerical integration of the subtracted double-real emission contribution for a 2-jet final state with a jet $p_T$ cut of 80 GeV and a centre-of-mass energy of $\sqrt{s}$=7 TeV. 
We observe that the numerical integration converges and for values of the phase space generation cut $y_{cut}=min\{s_{ij}/\hat{s}\} < 10^{-5}$ is independent of the value of the cut off.

\begin{table}
\centering
\caption{Partonic contributions to the $gg$ leading-$N_F$ contribution}
\label{tab-1}       
\begin{tabular}{lll}
\hline
NNLO contributions & perturbative order  \\\hline
$gg\to q\bar{q}gg$ & tree-level (RR) \\
$gg\to q\bar{q}g$   &  one-loop (RV)\\
$gg\to ggg$   & one-loop (RV)\\
$gg\to gg$    &  two-loop (VV)\\
$gg\to q\bar{q}$ & two-loop (VV)\\\hline
\end{tabular}
\end{table}

\begin{figure*}
\begin{minipage}[b]{0.5\linewidth}
\centering
\includegraphics[width=8cm,angle=0]{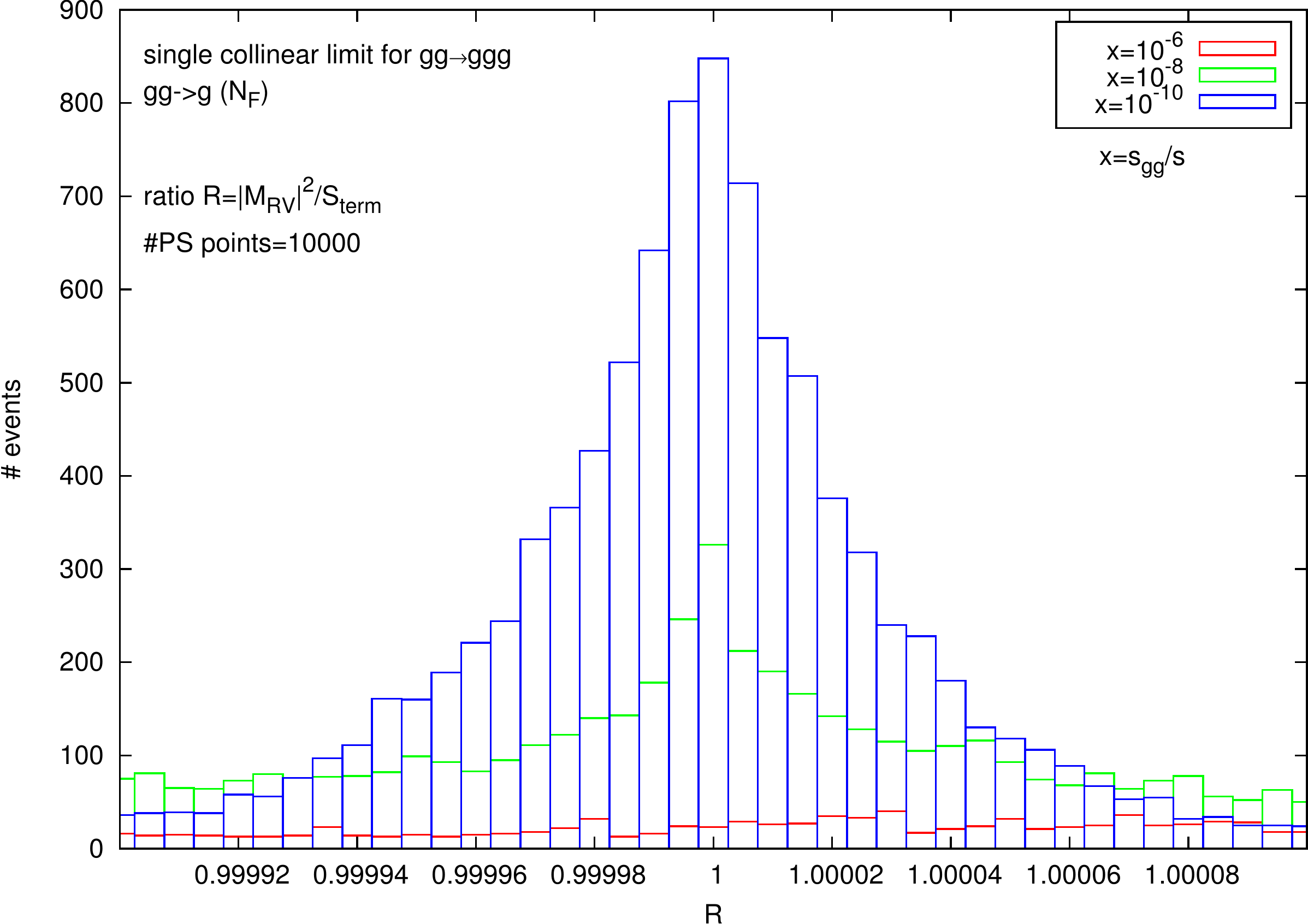}\\
\vspace{0.2cm}
(a)
\end{minipage}
\hspace{0.5cm}
\begin{minipage}[b]{0.5\linewidth}
\centering
\includegraphics[width=8cm]{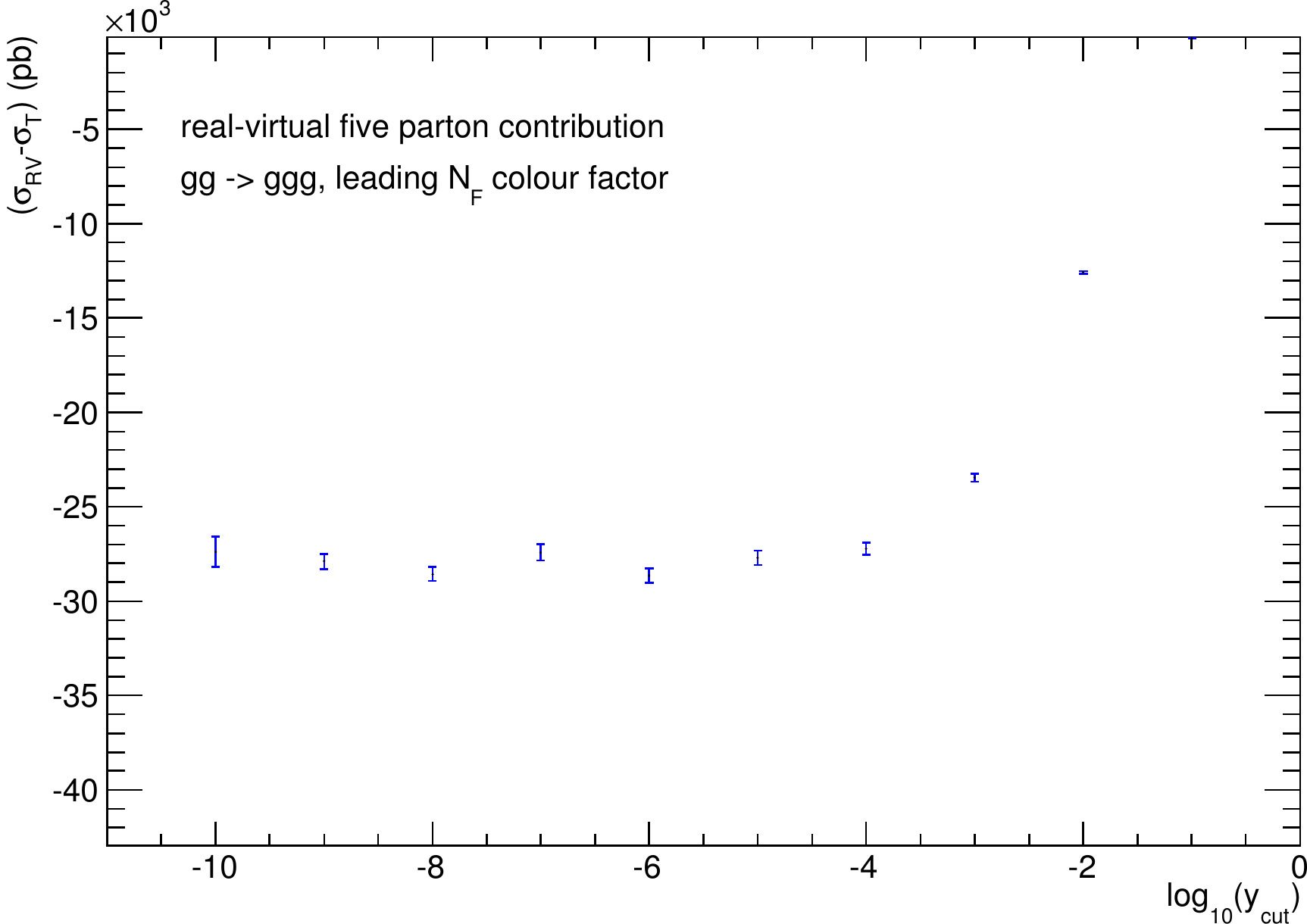}\\
\vspace{0.2cm}
(b)
\end{minipage}
\caption[]{Distribution of $R$ for 10000 phase space points for (a) single collinear $gg\to g$ configurations where the value of $x$ defined on the plot
controls the phase space approach to the singular configuration. On the right (b) the numerical values for the integration of the subtracted real-virtual emission contribution as a function of the phase space
generation cut-off parameter $y_{cut}$.}
\label{fig:RVint}
\end{figure*}

Upon integration of the single unresolved contributions in the double-real subtraction term, the RV contribution is explicitly IR-finite as all explicit poles in $1/\epsilon$ in the matrix element 
cancel in local and analytic form with the integrated subtraction terms arising from singly unresolved real emission, leaving behind a finite remainder that can be evaluated in 4 dimensions.
However the RV matrix elements can develop further implicit IR-singularities in the soft and collinear regions of the phase space of the RV contribution. For this reason
figure~\ref{fig:RVint} shows our numerical tests for the real-virtual emission process. As an example we show the generation of 10000 random phase space points for a
single collinear $gg\to g$ configuration to test the subtraction of the one-loop $N_{F}$ contribution to the $g\to gg$ splitting function. For this configuration we also produced 
the ratio plot by averaging over two phase space points related by a single $\pi/2$ rotation of the collinear system, as it was performed for the double real-emission
process. As for the double-real contribution we observed that the real-virtual subtraction term captures all the physical singularities of the matrix element of the process and that the numerical
integration of the subtracted real-virtual emission contribution converges and is independent of the value of the phase generation cut $y_{cut}$ for sufficiently small values of $y_{cut}$.

\section{Conclusions}
In this talk we briefly introduced and motivated the topic of QCD radiative corrections stressing the impact of including higher-order QCD effects in the theory predictions
and in particular we considered two-loop QCD corrections to jet production at hadron colliders. We presented the next-to-next-to-leading order (NNLO) QCD corrections to
dijet production in the purely gluonic channel retaining the full dependence on the number of colours combined with the NNLO leading-colour contribution in the $q\bar{q}$-channel. 

We considered also NNLO corrections to the $gg$-channel with quarks in the final state, namely the leading-$N_F$ double-real and real-virtual contributions. Upon integration
over the corresponding double and single unresolved antenna phase spaces the subtraction terms presented here contribute to the cancellation of the full IR singularities of the double virtual
two loop leading-$N_F$ corrections to $gg\to gg$ and $gg\to q\bar{q}$. We leave this exercise as well as the derivation of the updated predictions at NNLO in the $gg$-channel including the
leading $N_F$ contributions to future work.

\section*{Acknowledgements}
J.P. would like to thank James Currie for many useful and stimulating discussions and for comments on the manuscript. 
This research was supported by the Munich Institute for Astro-and Particle Physics (MIAPP) of the DFG cluster of excellence ``Origin and Structure of the Universe'', 
where J.P. was a workshop participant in July 2014. J.P. acknowledges support by an Italian PRIN2010 grant, and would like to thank the organisers and convenors
of the International Symposium on Multiparticle Dynamics (ISMD 2014) for their kind invitation and warm hospitality in Bologna.

%
%
%

\end{document}